\documentclass[12pt]{article}

\usepackage{rotating}
\usepackage{graphicx}
\usepackage{multirow}
\usepackage{multicol}
\usepackage{amsmath}
\usepackage[margin=2.3cm]{geometry}

\makeatletter

\usepackage[affil-it]{authblk}

\def\@author{}
\renewcommand\@author{\ifx\AB@affillist\AB@empty\AB@author\else
      \ifnum\value{affil}>\value{Maxaffil}\def\rlap##1{##1}%
    \AB@authlist \\*[0.3cm]on behalf of the CMS Collaboration \\[\affilsep]\AB@affillist
    \else  \AB@authors\fi\fi}
\makeatother

\begin{document}

\title{Search for the Standard Model Higgs boson produced by vector boson fusion and decaying to bottom quarks}

\author[1]{Giorgia Rauco%
  \thanks{Electronic address: \texttt{giorgia.rauco@cern.ch}}}
\affil[1]{Physik-Institut, Universit{\"a}t Z{\"u}rich, CH-8057 Z{\"u}rich, Switzerland}

\date{Dated: \today}

\setcounter{Maxaffil}{0}
\renewcommand\Affilfont{\itshape\small}

\maketitle

\begin{abstract}The search for the Standard Model (SM) Higgs boson (H) produced through the Vector Boson Fusion (VBF) mechanism and decaying to a pair of bottom quarks is reported. The used data have been collected with the CMS detector and correspond to an integrated luminosity of 19.8 fb$^{-1}$ of proton-proton collisions at €‰$\sqrt{s}=8$ TeV at the CERN LHC. Parked data have been exploited as well. This search resulted in an observed (expected) significance in these data samples for a H $\rightarrow$ b$\bar{\textnormal{b}}$ signal at a mass of 125 GeV of 2.2 (0.8) standard deviations. The cited signal strength,  $\mu=\sigma/\sigma_{\textrm{SM}}$, was measured to be 2.8$^{+1.6}_{-1.4}$. This result has been combined with other CMS searches for the SM Higgs boson decaying in a pair of bottom quarks exploiting other Higgs production mechanisms. The obtained combined signal strength is 1.0 $\pm$ 0.4, corresponding to an observed signal significance of 2.6 standard deviations for a Higgs boson mass of 125 GeV.
\end{abstract}

\section{Introduction}
In the Standard Model theory \cite{Glashow1961,Salam:1961en,Weinberg1967} the electroweak symmetry breaking is explained by the Brout-Englert-Higgs mechanism~\cite{PhysRevLett.13.321, Unnamed, PhysRevLett.13.585} which is responsible for the electroweak gauge bosons to acquire mass. This mechanism predicts the existence of a Higgs scalar boson. The observation of a new particle in the mass region around $125$~GeV, consistent with the Higgs boson, was announced by the ATLAS \cite{Aad2012} and CMS \cite{Chatrchyan12} experiments at CERN's {Large Hadron Collider} (LHC) on $4$ July 2012.

After the Higgs boson discovery, the main goal is now to precisely measure and study the properties of the recently discovered particle and to compare them with the ones expected for the predicted boson. One of the most interesting aspects to analyze is the coupling between the particle and the fermions and, in the SM theory, the most probable decay of the Higgs boson of m$_{\textnormal{H}} = 125$ GeV is in a pair of b-quarks. This process is particularly difficult to be observed at the LHC in the inclusive production (dominated by {Gluon Fusion}) since the QCD background is overwhelming. It is therefore searched in other production channels: {Vector Boson Fusion} (VBF), {Associated Vector Boson Production} (VH) and {Associated Top-Pair Production} (ttH), where the Higgs boson is produced in association with other particles, resulting in a more distinguishing signal topology. The VBF production channel is the one exploited in the analysis here presented. 

In the VBF process a quark of each one of the colliding protons radiates a W or Z boson that subsequently interact or fuse.

The two valence quarks are typically scattered away from the beam line and inside the detector acceptance, where they can be revealed as hadronic jets. The prominent signature of VBF is therefore the presence of two energetic hadronic jets, roughly in the forward and backward direction with respect to the proton beam line. As a result, the signal final state features are a central b-quark pair (from the Higgs decay) and a light- quark pair (u,d-type) from each of the colliding protons, in the forward and backward regions.

The overwhelmingly most relevant and irreducible background to the signal search comes from the QCD production of four jets events with true or mistagged b-jets. Other backgrounds arise from: (i) hadronic decays of Z or W bosons produced in association with additional jets, (ii) hadronic decays of top quark pairs, and (iii) hadronic decays of singly produced top quarks. The final expected signal yield includes also the contribution of the Higgs bosons produced in Gluon Fusion processes with at least two associated jets.

\section{Trigger}
The data used for this analysis were collected with two different trigger strategies:
\begin{enumerate}
\item  {\bf Dedicated VBF qqH $\mathbf{\rightarrow}$ qqb$\mathbf{\bar{\textnormal{{\bf b}}}}$ trigger}. 
A set of dedicated trigger paths was specifically designed and deployed for the VBF qqH $\rightarrow$ qqbb signal search, both for the L1 and HLT levels, and operated during the full 2012 data-taking period. This set of triggers, called nominal, collected the largest fraction of the signal event.

The L1 paths require the presence of three jets with p$_{\textrm{T}}$ above optimized thresholds X, Y, Z (X = 64 - 68 GeV, Y = 44 - 48 GeV, Z = 24 - 32 GeV) according to instantaneous luminosity. Among the three jets, at most one among the two p$_{\textrm{T}}$ leading jets can be in the forward pseudorapidity region, while the remaining two have to be central.
The HLT paths are seeded by the L1 paths described above, and require the presence of four jets with p$_{\textrm{T}}$ above thresholds that are again adjusted to the data-taking luminosity, p$_{\textrm{T}} >$ 75 - 82, 55 - 65, 35 - 48, and 20 - 35 GeV, respectively. At least one of the selected four jets must further fulfill minimum b-tagging requirements. To identify the two VBF-tagging jets two criteria have been exploited: (i) the pair with the smallest HLT b-tagging values; (ii) the pair with the maximum pseudorapidity opening. Both pairs are required to exceed variable minimum thresholds on $|\Delta\eta_{jj}|$ of 2.2-2.5, and of 200-240 GeV on the dijet invariant mass $m_{jj}$, depending on the instantaneous luminosity.

\item {\bf General-purpose VBF trigger.}
The L1 paths for the general-purpose VBF trigger require minimum hadronic activity in the event with a scalar  p$_{\textrm{T}}$ sum of 175 or 200 GeV, depending on the instantaneous luminosity. The HLT path is seeded by the L1 path described above, and requires the presence of at least two CaloJets with  p$_{\textrm{T}} >$ 35 GeV. Out of all the possible jet pairs in the event the pair with the highest invariant mass is selected as the most probable VBF tagging pair. The corresponding invariant mass m$_{jj}$ and absolute pseudorapidity difference $|\Delta\eta_{jj}|$ are required to be larger than 700 GeV and 3.5.
\end{enumerate}
The integrated luminosity collected with the first set of triggers was 19.8 fb$^{-1}$, while for the
second trigger it was 18.2 fb$^{-1}$ .
\section{Event reconstruction and selection}
The offline analysis uses reconstructed charged-particle tracks and candidates of the Particle-Flow (PF) algorithm \cite{CMS-PAS-PFT-10-001, CMS-PAS-PFT-1002, CMS-PAS-PFT-09-001}. Jets are reconstructed by clustering the PF candidates with the anti-k$_{\textnormal{T}}$ algorithm with distance parameter 0.5 and jets that are likely to be originated from the hadronization of b quarks are identified with the CSV b-tagger \cite{Chatrchyan:2012jua}.

The events used in the offline analysis are required to have at least four reconstructed jets and the four p$_{\textnormal{T}}$-leading ones are considered as the most probable b-jet and VBF jet candidates. A multivariate discriminant  taking into account the b-tag value, the b-tag ordering, the $\eta$ value, and the $\eta$ ordering is exploited to distinguish between the two jet types. 

The offline event selection is based upon the kinematic properties of the b-jet and VBF jets. Selected events are divided into two sets: {set A} and {set B}, whereof the selection requirements are shown in Table \ref{tab:sel}.

\begin{table}[h]
  \begin{center}
    \tabcolsep7pt\begin{tabular}{l|c|c}
      \hline
               & {set A}                  & {set B}   \\
     \hline
      {trigger}           & {dedicated VBF qqH ${\rightarrow}$ qqb$\rm{\bar{{{b}}}}$} & {general-purpose VBF trigger} \\ 
      \hline
      \multirow{2}{*}{} & \multirow{2}{*}{}                     &  p$_\textnormal{T,1,2,3,4}>30$GeV \\
        jets p$_{\textnormal{T}}$        & p$_\textnormal{T,1,2,3,4}>80,70,50,40$GeV &  \\
                        &                                       &  p$_\textnormal{T,1}+p_\textnormal{T,2}>160$GeV \\
      \hline 
      jets $|\eta|$     & $<4.5$                                & $<4.5$ \\
      \hline 
      b-tag             & at least 2 CSVL jets                  & at least 1 CSVM and 1 CSVL jets \\
      \hline
      $\Delta\phi_\textnormal{bb}$ & $<2.0$                           & $<2.0$ \\
      \hline
      \multirow{2}{*}{} & $m_\textnormal{jj}>250$GeV                 & $m_\textnormal{jj},\,m_\textnormal{jj}^\textnormal{trig}>700$GeV \\
        VBF topology    &                                       &  \\
                        & $|\Delta\eta_\textnormal{jj}| > 2.5$        & $|\Delta\eta_\textnormal{jj}|,\,|\Delta\eta_\textnormal{jj}^\textnormal{trig}|>3.5$ \\
      \hline    
     veto              & none                                  & events that belong to set A \\ 
   \hline 
    \end{tabular}
    
     \caption{Summary of selection requirements for the two analysis sets. Reprinted from \cite{PhysRevD.92.032008}.}\label{tab:sel}
  \end{center}
\end{table}

After all the selection requirements, $2.3\%$ of the VBF simulated signal events end up in set A and $0.8\%$ end up in set B. In set B $39\%$ of the signal events would also satisfy the requirements to enter set A. Such events are taken into set A and vetoed from set B, as noted in the last line of Table \ref{tab:sel}.
\section{Signal properties}
\subsection{Jet transverse-momentum regression}
In order to improve the b$\bar{\textnormal{b}}$ mass resolution a regression technique is applied. It is essentially a refined calibration for individual b-jets which takes into account the jet composition properties beyond the default jet-energy corrections. This regression technique mainly targets the b decays in a neutrino that lead to a substantial mismeasurement of the jet p$_{\textnormal{T}}$.

For this purpose a regression Boost Decision Tree (BDT), trained on simulated signal events, is applied. Its inputs include: (i) the jet p$_{\textnormal{T}}$, $\eta$ and mass; (ii) the jet-energy fractions carried by neutral hadrons and photons; (iii) the mass and the uncertainty on the decay length of the secondary vertex, when present; (iv) the event missing transverse energy and its azimuthal direction relative to the jet; (v) the total number of jet constituents; (vi) the p$_{\textnormal{T}}$ of the soft-lepton candidate inside the jet, when present, and its p$_{\textrm{T}}$ component perpendicular to the jet axis; (vii) the p$_{\textnormal{T}}$ of the leading track in the jet; (viii) the event's average p$_{\textrm{T}}$ density in the $y-\phi$ space.

The improvement on the jet p$_{\textnormal{T}}$ leads to an improvement on the dijet invariant mass resolution by approximately 17$\%$. 

\subsection{Discrimination between quark-and gluon-originated jets}
The VBF-tagging jets originate from the hadronization of a light (u,d-type) quark, while the jets produced in QCD processes are more likely to come from gluons. As a consequence, in order to further identify if the jet pair with the smallest b-tagging values among the four selected jets is a signal event or a background event, a quark-gluon discriminator \cite{Chatrchyan:2013jya, Charticyian2015, CMS-PAS-JME-13-002} is applied to the b-tag sorted jj candidate jets.\\
The discriminator exploits the differences in the showering and the fragmentation of gluons and quarks and it uses, as an input to a likelihood trained on gluon and quark jets from simulated QCD events, the following variables: (i) the jet constituents' major quadratic mean (RMS) in the $\eta-\phi$ plane; (ii) the jet constituents' minor quadratic mean (RMS) in the $\eta-\phi$ plane; (iii) the jet asymmetry pull (essentially a p$_{\mathrm{T}}$-weighted vector); (iv) the jet particle multiplicity; (v) the maximum energy fraction carried by a jet constituent.

\subsection{Soft QCD activity}

In the region between the two VBF-tagging jets (with the exception of the more centrally produced Higgs decay products), the QCD color flow is suppressed. In order to measure the additional hadronic activity associated with the main primary vertex, only charged tracks are used.

A collection of \emph{additional tracks} is built, selecting reconstructed tracks that: (i) have a  {\em high purity} quality flag; (ii) have p$_{\textnormal{T}}>$300 MeV; (iii) are not associated to any of the four leading jets; (iv) have a minimum longitudinal impact parameter, $|d_{z}(PV)|$ with respect to the event's main primary vertex; (v) satisfy $|d_{z}(PV)|<2$ mm and  $|d_{z}(PV)|<3\sigma_{z}(PV)$ where $\sigma_{z}(PV)$ is the uncertainty on $d_{z}(PV)$; (vi) are not in the region between the most b-tagged jets. This region is defined as an ellipse in the $\eta-\phi$ plane around the b-jets with axis $(a,b) = (\Delta R(bb)+1,1)$ where $\Delta R = \sqrt{(\Delta\eta_{bb})^2+(\Delta\phi_{bb})^2}$.

The additional tracks are then clustered in \emph{soft TrackJets} within the anti-$k_{\textnormal{T}}$ algorithm \cite{Cacciari:2008gp} (with R = 0.5). 

In order to discriminate between the signal and the QCD background, a discriminating variable $H_{T}^{soft}$ is used and it is defined as the scalar p$_{\textnormal{T}}$ sum of the soft TrackJets with p$_{\textnormal{T}}>$1 GeV. 

\section{Search for a Higgs boson}
In order to separate the overwhelmingly large QCD background from the Higgs boson signal, all the discriminating features have to be used in an optimal way. This is achieved by using a BDT multivariate discriminant, which exploits as input, variables very weakly correlated to the dynamics of the $\textnormal{b}\bar{\textnormal{b}}$ system, in particular to m$_{{\textnormal b}\bar{\textnormal{b}}}$. These variables are conceptually grouped into five groups: (i) the dynamics of the VBF jet system, expressed by $\Delta\eta_{\rm jj}$, $\Delta\phi_{\rm jj}$, and $ m_{\rm jj}$; (ii) the b jet content of the event, expressed by the CSV output for the two most b-tagged jets; (iii) the jet flavor of the event: quark-gluon likelihood (QGL) for all four jets; (iv) the soft activity, quantified by the scalar $p_{\textnormal{T}}$ sum $H_T^{\rm soft}$ of the additional ``soft'' TrackJets with $p_{\textnormal{T}}>1$ GeV, and the number $N^{\rm soft}$ of ``soft'' TrackJets with $p_{\textnormal{T}}>2$ GeV; (v) the angular dynamics of the production mechanism, expressed by the cosine of the angle between the $\rm jj$ and $\textnormal{b}\bar{\textnormal{b}}$ vectors in the center-of-mass frame of the four leading jets $\cos\theta_{\rm jj,bb}$.

Since the properties of the selected events are significantly different between the two selections (set A and set B) and two BDT's are trained. 

According to the BDT outputs, seven categories are defined: four for set A and three for set B. 

\begin{figure}[hbtp]
  \begin{center}
    \includegraphics[width=0.45\textwidth]{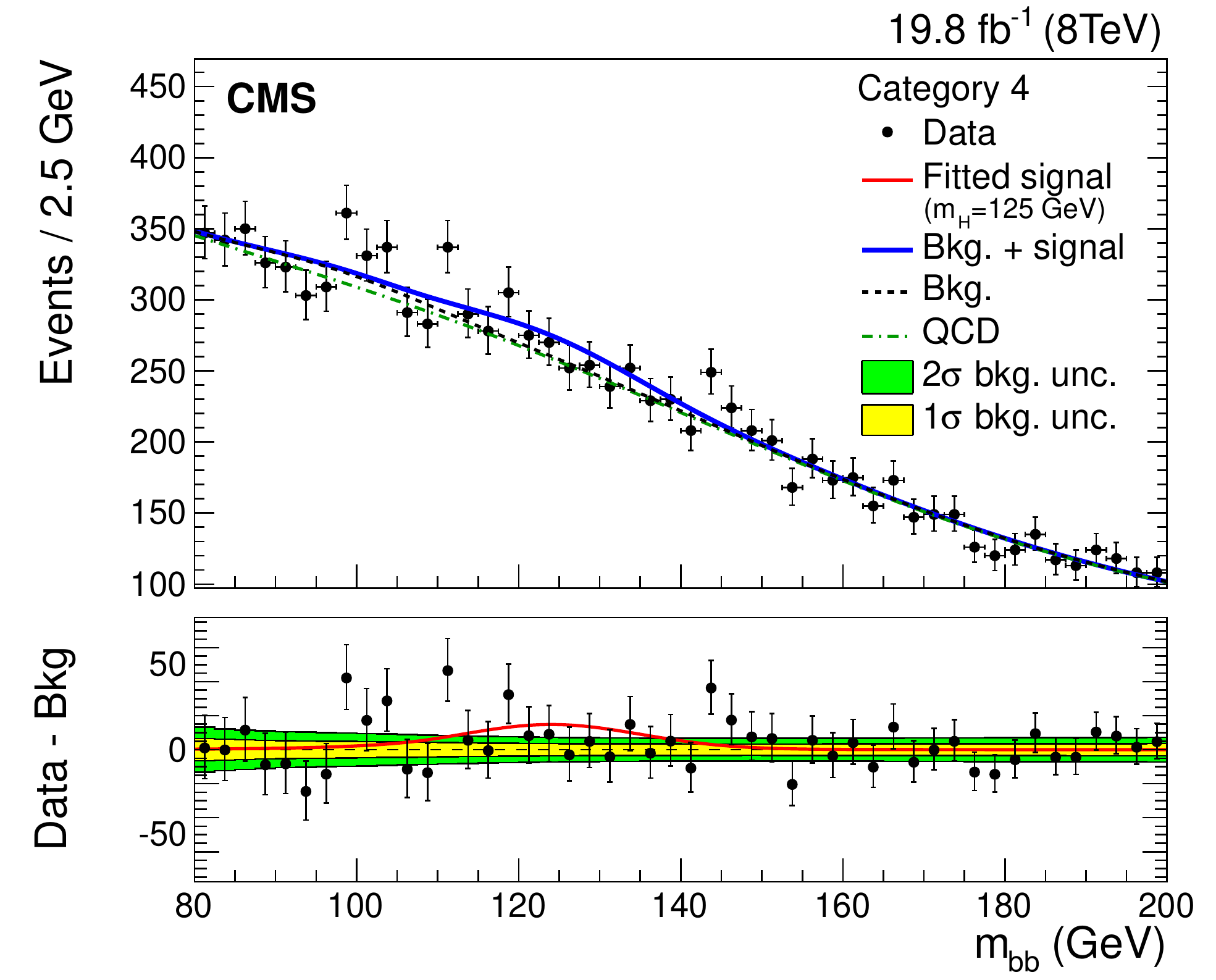}
    \includegraphics[width=0.45\textwidth]{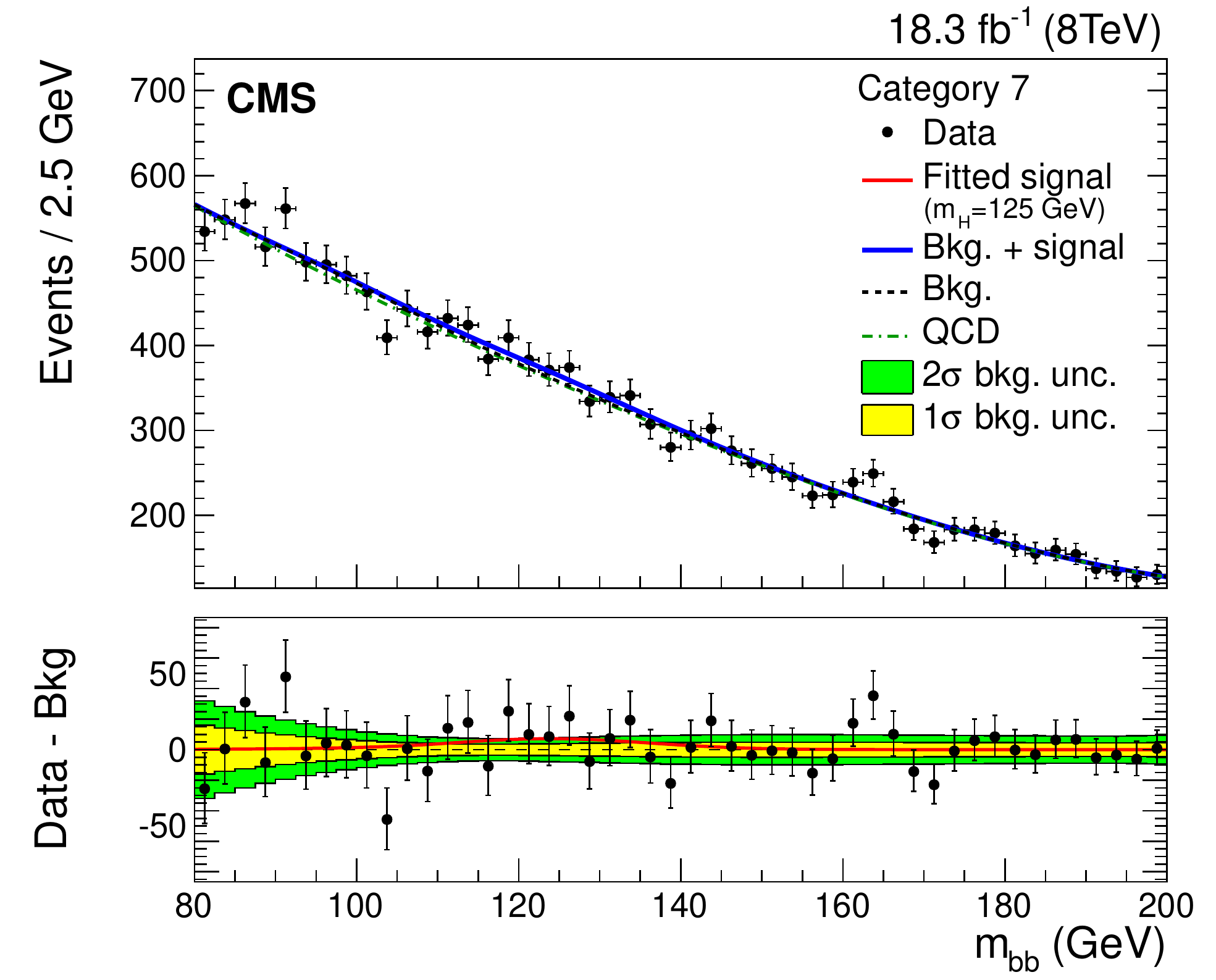}
    \caption{Fit for the Higgs boson signal (m$_\textnormal{H}=125$ GeV) on the invariant mass of the two b-jet candidates in the signal enriched event category of set A (left) and set B (right). Data is shown with markers. The solid line is the sum of the post-fit background and signal shapes, the dashed line is the background component, and the dashed-dotted line is the QCD component alone.  The bottom panel shows the background-subtracted distribution, overlaid with the fitted signal, and with the 1-$\sigma$ and 2-$\sigma$ background uncertainty bands. Reprinted from \cite{PhysRevD.92.032008}. }
    \label{fig:fitHiggs}
  \end{center}
\end{figure}

The QCD m$_{\textnormal{b}\bar{\textnormal{b}}}$ spectrum shape is assumed to be the same in all BDT categories of the same set of events. In reality small differences between the categories are present and to take into account this effect transfer functions are exploited (linear function in set A and quadratic in set B). With the introduction of the transfer functions, the fit model for the Higgs boson signal is
\begin{equation}\label{eq:fitH}
\begin{split}
 f_i(m_{\textnormal{b}\bar{\textnormal{b}}})=\mu_{\rm H}\cdot N_{i,\rm H}\cdot H_i(m_{\textnormal{b}\bar{\textnormal{b}}};k_{\rm JES},k_{\rm JER})+N_{i,\rm Z}\cdot Z_i(m_{\textnormal{b}\bar{\textnormal{b}}};k_{\rm JES},k_{\rm JER})+ \\
 N_{i,\rm Top}\cdot T_i(m_{\textnormal{b}\bar{\textnormal{b}}};k_{\rm JES},k_{\rm JER})+N_{i,\rm QCD}\cdot K_i(m_{\textnormal{b}\bar{\textnormal{b}}})\cdot B(m_{\textnormal{b}\bar{\textnormal{b}}};\vec{p}_{\rm set}),
\end{split}
\end{equation}
where the subscript $i$ denotes the category and $\mu_{\rm H},\,N_{i,\rm QCD}$ are free parameters for the signal strength and the QCD event yield. $N_{i,\rm H}$, $N_{i,\rm Z}$, and $N_{i,\rm Top}$ are the expected yields for the Higgs boson signal, the Z+jets, and the top background respectively. The shape of the top background $T_i(m_{\textnormal{b}\bar{\textnormal{b}}};k_{\rm JES},k_{\rm JER})$ is taken from the simulation (sum of the $\textnormal{t}\bar{\textnormal{t}}$ and single-top contributions) and is described by a broad gaussian. The Z/W+jets background $Z_i(m_{\textnormal{b}\bar{\textnormal{b}}};k_{\rm JES},k_{\rm JER})$ and the Higgs boson signal $H_i(m_{\textnormal{b}\bar{\textnormal{b}}};k_{\rm JES},k_{\rm JER})$ shapes are taken from the simulation and are parameterized as a crystal-ball function on top of a polynomial background. The position and the width of the gaussian core of the MC templates (signal and background) are allowed to vary by the free factors $k_{\rm JES}$ and $k_{\rm JER}$, respectively, which quantify any mismatch of the jet energy scale and resolution between data and simulation. Finally, the QCD shape is described by a Bernstein polynomial $B(m_{\textnormal{b}\bar{\textnormal{b}}};\vec{p}_{\rm set})$, common within the categories of each set, and whose parameters $\vec{p}_{\rm set}$ are determined by the fit, and a multiplicative transfer function $K_i(m_{\textnormal{b}\bar{\textnormal{b}}})$ that accounts for the shape differences between the categories. For set A, the Bernstein polynomial is of 5th order, while for set B it is of 4th order. Figure~\ref{fig:fitHiggs} show the simultaneously fitted m$_{\textnormal{b}\bar{\textnormal{b}}}$ distributions in the signal enriched categories for set A and set B, respectively.

\section{Results}
\begin{figure}[hbtp]
\begin{center}
    \includegraphics[width=0.6\textwidth]{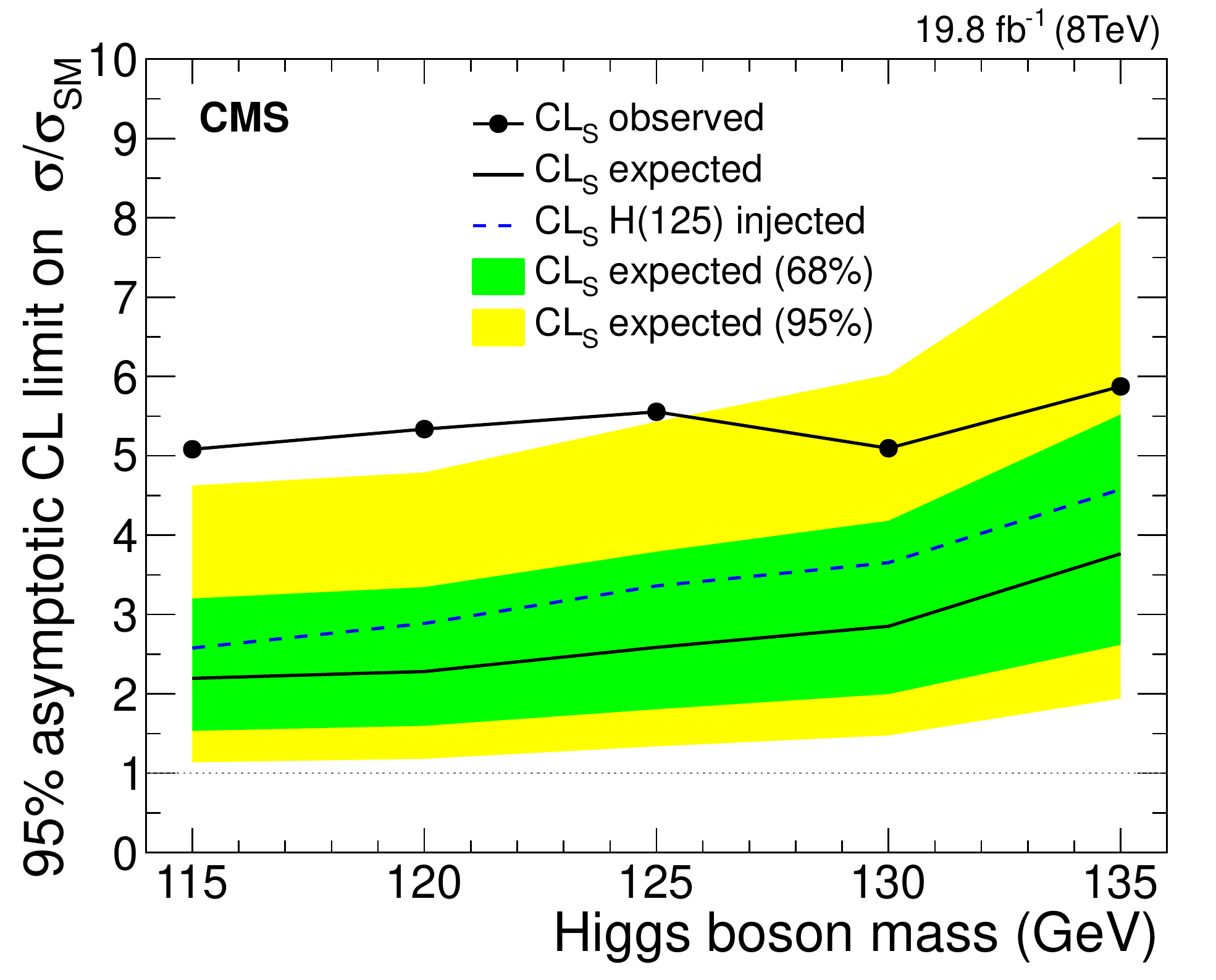} 
    \caption{Expected and observed $95\%$ confidence level limits on the signal cross section in units of the SM expected cross section, as a function of the Higgs boson mass, including all event categories. The limits expected in the presence of a SM Higgs boson with mass 125 GeV are indicated by the dotted curve. Reprinted from \cite{PhysRevD.92.032008}.}
    \label{fig:limits}
    \end{center}
\end{figure}

The models representing the two hypotheses, of background only, and of background+signal are fitted to the data, simultaneously in all the categories. The limits on the signal strength are computed with the Asymptotic CLs method~\cite{CLs}. Figure~\ref{fig:limits} shows the observed (expected) $95\%$ C.L. limit on the signal strength, as a function of the Higgs boson mass, which ranges from $5.1$ ($2.2$) at m$_\textnormal{H}=115$ GeV to $5.9$ ($3.8$) at m$_\textnormal{H}=135$ GeV, together with the expected limits in the presence of a SM Higgs boson with mass 125 GeV. For a 125 GeV Higgs boson signal the observed (expected) significance is 2.2 (0.8) standard deviations, and the fitted signal strength is $\mu=\sigma/\sigma_{\rm SM}=2.8^{+1.6}_{-1.4}$. The measured signal strength is compatible with the SM Higgs boson prediction $\mu=1$ at the $8\%$ level.

The search for the Standard Model Higgs boson decaying in a pair of bottom quarks, as described in the previous Sections, is performed by the CMS Collaboration in the VH \cite{vh}, VBF and ttH \cite{tth} production channels. The results of these searches are summarized in Table~\ref{tab:summary}, along with the resulting combined results.
\begin{table}[h]
\begin{center}
 \tabcolsep7pt\begin{tabular}{c|c|cc|cc}
 \hline
	{H $\rightarrow$ b${\bar{\textnormal{{b}}}}$} & {Best-fit (${68\%}$ CL)} & \multicolumn{2}{c|}{{Upper Limits (${95\%}$ CL)}} & \multicolumn{2}{c}{{Signal significance}} \\
	{channel} & {Observed} & {Observed} & {Expected} & {Observed} & {Expected} \\\hline
    VH & 0.89 $\pm$ 0.43 & 1.68 & 0.85 & 2.08 & 2.52 \\
    ttH & 0.7 $\pm$ 1.8 & 4.1 & 3.5 & 0.37 & 0.58 \\
    VBF & 2.8$^{+1.6}_{-1.4}$ & 5.5 & 2.5 & 2.20 & 0.83 \\ \hline 
    combined & 1.03$^{+0.44}_{-0.42}$ & 1.77 & 0.78 & 2.56 & 2.70 \\
    \hline
    \end{tabular}
 \caption{Observed and expected 95$\%$CL limits, best fit values and significance on the signal strength parameter $\mu = \sigma/\sigma_{\textnormal{SM}}$ at m$_{\textnormal{H}}$ = 125 GeV, for each H $\rightarrow$ b$\bar{\textnormal{b}}$ channel and combined. Reprinted from {\cite{PhysRevD.92.032008}}.}\label{tab:summary}
  \end{center}
\end{table}
 
\clearpage,


\end{document}